\documentclass[twocolumn,letter]{jpsj2} %% two-column layout
\title{
Unscreening Effect on Fe-Pnictide Superconductor
}

\author{Yuki \textsc{Fuseya}$^{1}$
\thanks{E-mail: fuseya@hosi.phys.s.u-tokyo.ac.jp},
Toshikaze \textsc{Kariyado}$^{1}$
, and 
Masao \textsc{Ogata}$^{1, 2}$ 
}

\inst{
$^{1}$ Department of Physics, University of Tokyo, 7-3-1 Hongo, Bunkyo-ku, Tokyo 113-0033 \\
$^{2}$ JST, TRIP, 5 Sanbancho, Chiyoda, Tokyo 102-0075
}

\abst{
	We study a microscopic mechanism of Fe-pnictide superconductor, considering the screening effects of Coulomb interaction in addition to the conventional spin-fluctuation mechanism.
	It is shown that, by electron doping, the transition temperature of superconductivity increases due to the ``unscreening" effect even though the density of states decrease, while that of spin-density wave rapidly decreases due to breaking of nesting conditions.
	Our results give a clear interpretation to the mystery of interrelation between $T_{\rm c}$ and the density of states observed in the Fe-pnictide superconductors.
}

\kword{Fe-pnictide superconductor, screening, extended $s$-wave pairing, semimetal superconductor, excitonic phase}

\usepackage{mathrsfs}

\newcommand{\bk}{\boldsymbol{k}}
\newcommand{\bq}{\boldsymbol{q}}
\newcommand{\bQ}{\boldsymbol{Q}}

\newcommand{\Tc}{T_{\rm c}}
\newcommand{\EF}{E_{\rm F}}
\newcommand{\qs}{q_{\rm s}}
\newcommand{\Eg}{E_{\rm g}}

\begin{document}
\maketitle

%\section{Introduction} %% No sections necessary for express letters, letters and short notes
	%
	Recent discovery of superconductivity (SC) at high temperature in iron-pnictide family of compounds\cite{Kamihara08,Takahashi08,XHChen08,GFChen08,Ren08} raises the question whether a new mechanism is responsible for the SC.
	The parent compound LaFeAsO is metallic but shows an anomaly near 150K in both resistivity and d.c. magnetic susceptibility\cite{Kamihara08}, suggesting a phase transition to a spin-density-wave (SDW) phase\cite{Cruz08}.
	By doping of F ions into O sites, corresponding to electron doping, this SDW phase is suppressed and then LaFeAsO$_{1-x}$F$_x$ undergoes superconducting transition with the highest transition temperature, $\Tc$, of $\sim 26$K at 5-11 atom \%.
	The transition temperature increases by applying pressure\cite{Takahashi08} or replacement of La with other rare earth elements\cite{XHChen08,GFChen08,Ren08}.

	The electronic structure of normal state of LaFeAsO$_{1-x}$F$_{x}$ has been revealed by first principle calculations as follows\cite{Singh08,Haule08,Mazin08,Cao08,Kuroki08,Ishibashi08,Yin08,Ma08}.
	The calculated density of states, $N(E)$, exhibits van Hove singularities above and below the Fermi energy, $\EF$\cite{Singh08,Haule08,Ishibashi08}.
	$N(E)$ rapidly changes near $\EF$ with a negative gradient, $d N(E)/dE |_{E=\EF}<0$, so that $N(\EF)$ rapidly decreases with electron doping\cite{Ma08}.
	The Fermi surfaces for undoped LaFeAsO consist of two high-velocity electron cylinders around the zone edge $M$, two low-velocity hole cylinders around the zone center $\Gamma$, plus a heavy three-dimensional (3D) hole pocket centered at $Z$\cite{Singh08,Ma08}.
	The 3D hole pocket soon disappears with electron doping, leaving the two electron and hole cylinders, i.e., almost two-dimensional (2D) systems\cite{Mazin08,Ma08}.

	There have been several mechanisms proposed for the SC of Fe-pnictides\cite{Kuroki08,Baskaran08,Lee08,Barzykin08,Nomura08,Yanagi08,Ikeda08}.
	Theories based on the detailed band-structure, where the tight-binding dispersion (more than five bands) is obtained by fitting the first principle calculations, have reported that the extended $s$-wave singlet pairing (called as $s_{\pm})$ is the most stable\cite{Kuroki08,Nomura08,Yanagi08,Ikeda08}.
	The SC gaps fully open on the Fermi surfaces, and the SC order parameters change their sign between the hole- and the electron-pockets.
	This SC state mainly originates from the nesting between the electron- and hole-pockets, which generates the SDW at zero doping.

	Although these theoretical analysis seems to be reasonable in explaining several experiments, there are some experiments which are difficult to understood.
	The photoemission spectroscopy shows that the intensity of the spectra at $\EF$ decreases with electron doping, as is expected from the band calculations.
	Nevertheless, $\Tc$ does not decrease with this electron doping.
	Moreover, the intensity for LaFeAsO$_{0.94}$F$_{0.06}$ is smaller than that for LaFePO$_{0.94}$F$_{0.06}$, though $\Tc$ of LaFeAsO$_{0.94}$F$_{0.06}$ is much higher.
	These experimental results strongly suggest that $\Tc$ tends to increase when $N(\EF)$ decreases, which is opposite from the conventional theories of SC.
	The nuclear magnetic relaxation rate, $1/T_1$, indicates that $1/T_1 T$ (or the spin fluctuation) is rapidly suppressed with electron doping ($x=$0.04-0.11), while $\Tc$ is hardly affected.
	It is also difficult to explain this result on the basis of the spin-fluctuation mechanism.

	This Letter is intended to give a possible solution for these mysterious behaviors, which will be essential for the superconducting mechanism of Fe-pnictides.
	We shall introduce an idea of ``unscreening effect"; as $N(\EF)$ decreases, the screening effect is weakened, so that the Coulomb interaction become long-ranged.
	We show that this unscreened effect actually increases $\Tc$, and causes the apparently independent behavior of SC from SDW.

%%%%%%%%%%%%%%%%%%%%%%%%%%%%%%%%%%%%%%%%%%%%%%%%%%%%%%%%%%%%%%%%%%%%%%%%
%\section{Theory}
%-----------------------------------------------------------------------
	%
%=======================================================================
\begin{figure}[bt]
\begin{center}
\includegraphics[width=7cm]{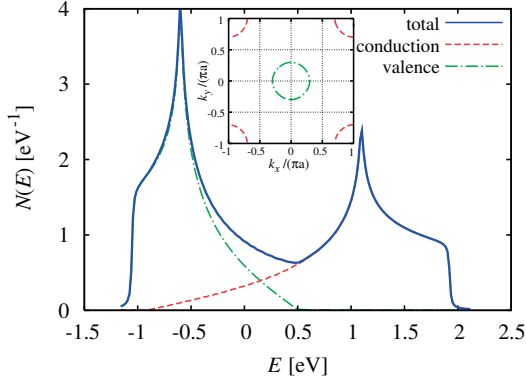}
\end{center}
\caption{
(Color online) Density of states, $N(E)$, of the two-band model.
The inset shows the Fermi surfaces for undoped ($\bar{\mu}=0.2045$ eV) case.
}
\label{DOS}
\end{figure}
%=======================================================================
	%
	First, we construct an effective model for the Fe-pnictide.
	We adopt the following 2D two-band model in order to make our arguments as clear and transparent as possible.
	%
	%The actual band structure of Fe-pnictide consists of two hole- and two electron-pockets as explained above.
	%
	The key point of the present work is the property of $N(E)$, so that we introduce a model which reproduces $N(E)$ obtained by the first principle calculation of LaFeAsO\cite{Singh08,Haule08,Ishibashi08}.
	The electronic dispersion of our two-band model is given by
\begin{align}
	\xi_{\ell \bk} = 
	\alpha \left[-2t_\ell \left( \cos K_x + \cos K_y \right)
	+4t_\ell
	\right]^{1/2} + E_g/2 -\bar{\mu},
	\label{two-band}
\end{align}
	where $K_i = k_i a  -Q_i$, $\alpha=1$ for the conduction band ($\ell={\rm c}$), and $K_i = k_i a$, $\alpha=-1$ for the valence band ($\ell={\rm v}$) with $\bQ = (\pi, \pi)$, which corresponds to the $M$-point.
	We took parameters as $t_{\rm c}=1$ eV, $t_{\rm v}=0.3$ eV, $E_g=-1.4$ eV, $a =4.03552$ \AA\cite{Kamihara08}.
	$N(E)$ obtained in this two-band model is shown in Fig. \ref{DOS}, which reproduces the results of the first-principle calculations\cite{Ishibashi08,Singh08,Haule08}.
	The carrier densities of conduction and valence band are equal, when the chemical potential is $\bar{\mu}=0.2045$ eV ($\equiv \mu_{\rm 0}$).
	Hereafter, we measure the chemical potential from the undoped case as $\mu=\bar{\mu}-\mu_0$.
	The inset of Fig. \ref{DOS} shows the Fermi surfaces for $\mu=0$.
	There are van Hove singularities above and below ($E \simeq -0.6, 1.1$ eV) the Fermi level and the gradient of $N(E)$ is negative near $\mu=0$.
	$N(E)$ takes its minimum value above the Fermi level ($E \simeq 0.5$ eV), where the hole carries disappear.
	%
	%These properties agree well with the first principle calculations\cite{Ishibashi08,Singh08,Haule08}.
	%
	
	With this dispersion, we consider the following model Hamiltonian
\begin{align}
	\mathscr{H}&= \sum_{\ell\bk\sigma}
	\xi_{\ell \bk}c_{\ell \bk \sigma}^\dagger c_{\ell \bk \sigma}
	+\frac{1}{2}
	\sum_{\ell \ell'\bk \bq}
	V (\bq) 
	c_{\ell \bk+\bq \sigma}^\dagger
	c_{\ell' \bk-\bq \sigma'}^\dagger
	c_{\ell' \bk \sigma'}
	c_{\ell \bk \sigma},
\end{align}
	where $c_{\ell \bk \sigma}$ is the annihilation operator of the electron of the momentum $\bk$ and the spin $\sigma$ belonging to the band $\ell$.
	$V(\bq)$ is the Fourier transform of the Coulomb interaction.
	Here we adopt the 2D form given by $V (\bq)=2\pi e^2/\left[\epsilon(\bq) |\bq|\right]$\cite{Takada78,Ando82}, where $\epsilon (\bq)$ is the dielectric function, for the theoretical consistency with the 2D two-band model of eq. (\ref{two-band}).
	Note that when we adopt the 3D form given by $V (\bq)=4\pi e^2/\left[\epsilon(\bq) q^2\right]$, the present unscreening effect is expected to be also valid and more remarkable due to the $q^2$-factor.
	One usually approximates $\epsilon (\bq) \simeq 1+\qs/|\bq|$, where $\qs=2\pi e^2 N(\EF)$ being the inverse of the screening radius (the Thomas-Fermi approximation).
	Then, the Coulomb interaction is given as 
\begin{align} 
	V (\bq)=\frac{2\pi e^2}{ |\bq| + \qs }.
	\label{Coulomb}
\end{align}
	Obviously, when $N(\EF)$ decreases, $V(\bq)$ is strengthened, i.e., the screening is weakened due to small $N(\EF)$; we call this ``unscreening" effects.
	Such unscreening effects have been investigated in the studies of the excitonic phase\cite{Kozlov65,Halperin68}.
	The transition temperature of the excitonic phase takes the maximum when the conduction band touches the valence band (e.g., $|\Eg|\to0$, in the present model), namely, $N(\EF)$ vanishes.
	This is due to the unscreening effect.
	Note that the triplet exciton pairing is equivalent to the SDW order parameter between electron- and hole-pocket\cite{Halperin68}, which exactly corresponds to the case in the Fe-pnictides\cite{Barzykin08}.
	%
	%Actually, the first correction to the SC pairing for the present case is the excitation of the excitonic pairing.

%=======================================================================
\begin{figure}[tb]
\begin{center}
\includegraphics[width=8cm]{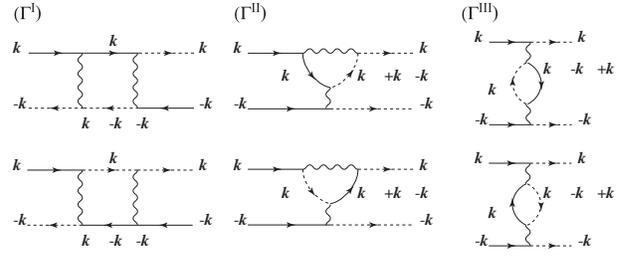}
\end{center}
\caption{
Diagrammatic representations for the first correction to the pairing interaction, $\Gamma_{\ell \ell'}^{(1)}$.
Solid and dashed lines denote the band $\ell$ and $\ell'$.
}
\label{diagrams}
\end{figure}
%=======================================================================
	%
	
	%Since the unscreening effect strengthen the Coulomb interactions, 
	We next verify that the unscreening effect actually increases $\Tc$.
	%
	%We also study the competition between SC and SDW, since the unscreening effect can affect both SC and SDW.
	%
	For an estimation of $\Tc$, we solve the linearlized gap equation for two-band,
\begin{align}
	\lambda_{\rm sc} \Delta_{\ell \bk} =
	-\sum_{k'}\Gamma_{\ell \ell'} (\bk, \bk') 
	\frac{\Delta_{\ell' \bk'}}{2\xi_{\ell' \bk'}}
	\tanh\frac{\xi_{\ell' \bk'}}{2T},
	\label{gap eq}
\end{align}
	%
	%
	%First we consider the Coulomb interaction keeping its momentum dependence.
	%
	where the pairing interaction, $\Gamma_{\ell \ell'}$, are given for the first corrections (Fig. \ref{diagrams}) as:
%	$\Gamma_{\ell \ell'}^{(1)}(\bk, \bk')
%	= 2\Gamma_{\ell \ell'}^{\rm I} (\bk, \bk')
%	+4\Gamma_{\ell \ell'}^{\rm II} (\bk, \bk')
%	-4\Gamma_{\ell \ell' }^{\rm III} (\bk, \bk')$,
\begin{align}
	\Gamma_{\ell \ell'}^{(1)}(\bk, \bk')
	&= 2\Gamma_{\ell \ell'}^{\rm I} (\bk, \bk')
	+4\Gamma_{\ell \ell'}^{\rm II} (\bk, \bk')
	-4\Gamma_{\ell \ell' }^{\rm III} (\bk, \bk'), \\
	%I
	\Gamma_{\ell \ell' }^{\rm I} (\bk, \bk') &= \sum_{\bk''}V(\bk''-\bk)V(\bk'-\bk'') \chi_{\ell \ell'0}(\bk+\bk', \bk''), \\
	%II
	\Gamma_{\ell \ell' }^{\rm II} (\bk, \bk') &= \sum_{\bk''}V(\bk''-\bk)V(\bk'-\bk) \chi_{\ell \ell'0}(\bk-\bk', \bk''), \\
	%III
	\Gamma_{\ell \ell'}^{\rm III} (\bk, \bk') &= -\left[V(\bk'-\bk)\right]^2 \sum_{\bk''}\chi_{\ell \ell'0}(\bk-\bk', \bk''), 
\end{align}
	with $\chi_{\ell \ell'0}(\bq, \bk)=\left( f_{\ell' \bk-\bq} - f_{\ell \bk} \right)/ \left( \xi_{\ell \bk} - \xi_{\ell' \bk-\bq}\right)$ and $f_{\bk}$ being the Fermi distribution function.
	$\chi_{{\rm cv}0 (\bq, \bk)}$ is the excitation of the excitonic pairing.
	%
%%=======================================================================
%\begin{figure}[tb]
%\begin{center}
%\includegraphics[width=7cm]{vq2d.eps}
%\end{center}
%\caption{Basically, figures and tables must be located near the place where they appear for the first time in the text.}
%\label{Gamma}
%\end{figure}
%%=======================================================================
	%
	%The results of $\Gamma_{\rm c v}^{\rm I\sim III}(\bq, 0)$ are shown in Fig. \ref{Gamma} for $\mu=0$ and $T=312.5$ K. %, and $\epsilon=3$.
	%
	%For $\mu=0$, each $\Gamma_{\rm cv}$ exhibits peak structure at $\bq=(\pi, \pi)$, reflecting the nesting between the conduction and valence band.
	%
	%These peaks are reduced and slightly shifted away from $\bq=(\pi, \pi)$ by doping.
	%
	%In $\Gamma_{\rm cv}^{\rm I}$, there is no peak structure around $\bq=0$, since the long wave-length component of $V(\bq)$ are integrated out.
	%
	%On the other hand, in $\Gamma_{\rm cv}^{\rm III}$, there is a peak at $\bq=0$ since the $V(\bq)$ is not integrated.
	%
	%$\Gamma_{\rm cv}^{\rm II}$ also exhibits a small peak at $\bq=0$.
	%
	%These peaks are solid against the doping.

%%%%%%%%%%%%%%%%%%%%%%%%%%%%%%%%%%%%%%%%%%%%%%%%%%%%%%%%%%%%%%%%%%%%%%%%
%\section{Results}
%-----------------------------------------------------------------------
%=======================================================================
\begin{figure}[tb]
\begin{center}
\includegraphics[width=7cm]{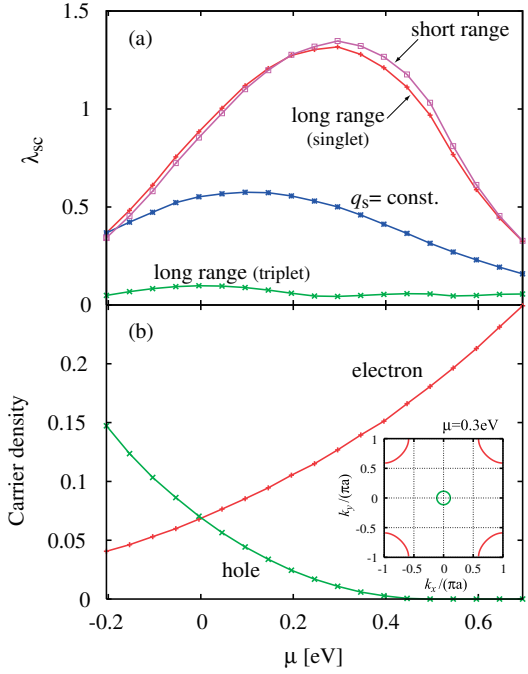}
\end{center}
\caption{
(Color online)
(a) Eigenvalues of the gap equation, $\lambda_{\rm sc}$, with the long range $V(\bq)$ (both for singlet and triplet), the constant $q_{\rm s}$, and the short range $V_0 (\epsilon =1.15)$.
(b) Carrier densities for electrons and holes.
The inset shows the Fermi surfaces for $\mu=0.3$ eV.
}
\label{lambda}
\end{figure}
%=======================================================================

	%
	The eigenvalue calculated with $\Gamma_{\ell \ell'}^{(1)}(\bq)$ is shown by a curve with ``long range" in Fig. \ref{lambda} (a).
	The calculations are carried out for the system of $32\times 32$ meshes in the Brillouine zone.
	The eigenvalue $\lambda_{\rm sc}$ increases by doping electrons even though $N(\EF)$ decreases (cf. Fig. \ref{DOS}).
	This behavior is opposite to the conventional BCS mechanism, where $\Tc$ is proportional to $\exp \left[-1/VN(E_F) \right]$.
	This anomalous increase is due to the unscreening effect.
	$\lambda_{\rm sc}$ takes its maximum at around $\mu=0.3$ eV.
	This is rather surprising, since the hole pocket is tiny in this case and only the electron pocket is sizable (Fig. \ref{lambda}(b)), i.e., there is no nesting property.
	In such a situation, it is difficult for the spin-fluctuation mechanism to develop the superconducting instability.
	These features are peculiar to the present unscreening mechanism, suggesting that the band located near the $\EF$ can contribute to the superconducting mechanism even if it does not appear as a visible Fermi surface.
	We also find that the gap functions are $s_{\pm}$-wave, i.e., they are fully gapped and almost constant on the Fermi surface changing their sign between different bands, ${\rm sign} (\Delta_{\rm c})=-{\rm sign} (\Delta_{\rm v})$.
	The magnitude of each gap varies with respect to $\mu$ as, roughly speaking, 
\begin{align}
	\Delta_{\rm c}^2 / \Delta_{\rm v}^2
	\simeq 
	N_{\rm v}(\EF) / N_{\rm c}(\EF).
\end{align}
	Note that the $\lambda_{\rm sc}$ for the odd-parity triplet pairing is vanishingly small for whole region.

	If we assume that $\qs$ is independent from $\mu$, $\lambda_{\rm sc}$ has a maximum at around $\mu=0$ and keeps monotonically decreasing with increasing $\mu$ as shown by a curve with ``$q_{\rm s}={\rm const.}$" in Fig. \ref{lambda} (a).
	In this case the property of $\lambda_{\rm sc}$ is naturally understood within the conventional spin-fluctuation mediated SC.
	From this, we conclude that the behavior that $\lambda_{\rm sc}$ takes a maximum away from $\mu=0$ is due to the $\mu$-dependence of $\qs$, i.e., the unscreening effect.

	For comparison, we also study a case in which $V(\bq)$ is approximated as a $\bq$-independent $V_0=2\pi e^2/\epsilon \qs$ with $\epsilon$ being the dielectric constant.
	This Coulomb interaction is short ranged and its magnitude depends on $\mu$ through $q_{\rm s}$.
	%
	%It is often useful for the qualitative discussions to approximate the Coulomb interaction to be of short range and the magnitude is of the order of $V_0=2\pi e^2/\epsilon \qs$, where $\qs$ depends on $\mu$ and $\epsilon$ being the dielectric constant.
	%
	%Within this approximation, the contribution from $\Gamma^{\rm II}$ and $\Gamma^{\rm III}$ compensate each other, so that only $\Gamma^{\rm I}$ remains.
	%
	The result with $V_0 (\epsilon=1.15)$ is shown in Fig. \ref{lambda} (a) denoted by ``short range".
	The behavior of $\lambda_{\rm sc}$ with $V_0$ agrees quite well with that obtained with the momentum-dependent Coulomb interaction, $V(\bq)$.

	The unscreening effect can affect both SC and SDW.
	So here we study the competition between SC and SDW, using the $\bq$-independent Coulomb interaction, $V_0$, since it reproduces $\lambda_{\rm sc}$ obtained by $V(\bq)$.
	We carry out the random phase approximation (RPA) for the pairing interaction in eq. (\ref{gap eq}) as follows:
	%
	%Then, we can carry out the random phase approximation (RPA) with $V_0$ as follows:
	%
\begin{align}
	\Gamma_{\ell \ell'}^{\rm RPA} (\bk, \bk') &= V_0
	+\frac{2V_0^2\tilde{\chi}_{\ell \ell' 0}(\bk+\bk')}
	{1-V_0\tilde{\chi}_{\ell \ell' 0}(\bk+\bk')}
	+\frac{4V_0^2\tilde{\chi}_{\ell \ell' 0}(\bk-\bk')}
	{1-V_0\tilde{\chi}_{\ell \ell' 0}(\bk-\bk')}
	\nonumber\\&
	-\frac{4V_0^2\tilde{\chi}_{\ell \ell' 0}(\bk-\bk')}
	{1+2V_0\tilde{\chi}_{\ell \ell' 0}(\bk-\bk')},
\end{align}
	where $\tilde{\chi}(\bq)=\sum_{\bk}\chi(\bq,\bk)$.
	For the transition temperature of SDW, we obtained the following gap equation within the parallel approximation to that of SC,
\begin{align}
	\lambda_{\rm sdw} \Delta_{\bk}'&= \sum_{\bk'} \Gamma_{\rm s}(\bk, \bk') 
	\frac{\Delta_{\bk'}'}
	{\xi_{{\rm v}\bk}-\xi_{{\rm c}\bk-\bQ}}
	\nonumber\\&\times
	\frac{1}{2}
	\left\{
	\tanh \frac{\xi_{{\rm v}\bk}}{2T}
	-\tanh \frac{\xi_{{\rm c}\bk-\bQ}}{2T}
	\right\},
	\\
	\Gamma_{\rm s}(\bk, \bk')
	&=V_0 
	+\frac{4V_0^2\tilde{\chi}_{\ell \ell' 0}(\bk-\bk')}
	{1-V_0\tilde{\chi}_{\ell \ell' 0}(\bk-\bk')}
	%
	%\nonumber\\&
	-\frac{4V_0^2\tilde{\chi}_{\ell \ell' 0}(\bk-\bk')}
	{1+2V_0\tilde{\chi}_{\ell \ell' 0}(\bk-\bk')},
\end{align}
	where $\lambda_{\rm sdw}$ and $\Delta'$ are the eigenvalues and gap function of SDW, respectively.
	%
	%The global property of $\lambda_{\rm sc}$ is the same as that obtained with $V(\bq)$, and the absolute value is enhanced due to the higher-order corrections.
	%
%=======================================================================
\begin{figure}[tb]
\begin{center}
%\begin{flushleft}
%(a)
%\end{flushleft}
%
%\vspace{-3.5zw}
%\includegraphics[width=7cm]{mu-t_ex_blue.eps}
%\begin{flushleft}
%(b)
%\end{flushleft}
%
%\vspace{-4zw}
%\includegraphics[width=7cm]{mu-t_sc_hot.eps}
\includegraphics[width=8cm]{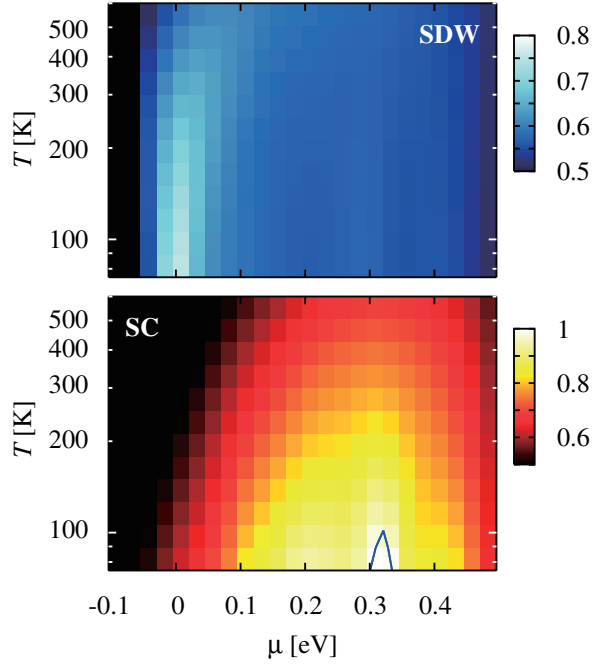}
\end{center}
\caption{
(Color online) Contour plots of $\lambda_{\rm sdw}$ (upper panel) and $\lambda_{\rm sc}$ (lower panel) obtained by the RPA with $V_0 (\epsilon = 2.8)$.
The solid line indicates $\lambda_{\rm sc}=1$.
}
\label{RPA}
\end{figure}
%=======================================================================
	%
	
	The results of these RPA are shown in Fig. \ref{RPA} setting $\epsilon=2.8$\cite{dielectric,Nakamura08}.
	It is clearly seen that the SDW instability is the largest at around $\mu=0$, whereas the SC one is the largest at around $\mu=0.3$ eV.
	This means that, for SDW, the effect of the nesting condition is more relevant than the unscreening effect.
	As a result, SDW is realized around the undoped case, $\mu=0$, and is rapidly suppressed by electron doping, while $s_{\pm}$-wave superconductivity is replaced.
	%

%%%%%%%%%%%%%%%%%%%%%%%%%%%%%%%%%%%%%%%%%%%%%%%%%%%%%%%%%%%%%%%%%%%%%%%%
%\section{Discussions}
%-----------------------------------------------------------------------
	
	Let us discuss implications of  the present results to the experiments on LaFeAsO$_{1-x}$F$_x$.
	The behavior that $\Tc$ increases as $N(\EF)$ (or the intensity of photoemission spectra) decreases can be naturally understood from the present unscreening mechanism. %, where the Coulomb interaction is strengthened due to low carrier density.
	%
	%The SDW instability is much sensitive to the nesting condition than to the unscreening effect, leading to the rapid suppression of the SDW phase.
	%
	%Our results indicate that the unscreening effect is much effective in superconductivity than in SDW, which is sensitive to the nesting condition.
	%
	In the case of LaFeAsO$_{1-x}$F$_x$, the nesting conditions is best at $x=0$ (undoped), since the carrier densities of electrons and holes are equal. %, so that the almost circle electron- and hole-pokes are well nested.
	On the other hand, the unscreening effect will be most effective when the hole-pockets almost vanish, whose doping ratio is estimated to be $x=$0.3-0.4\cite{doping}.
	These two characteristic features are well separated in the doping axis $x$ for LaFeAsO$_{1-x}$F$_x$, resulting in the apparent independence of $\Tc$ on the spin fluctuation estimated from $1/T_1 T$.
	%

%=======================================================================
\begin{figure}[tb]
\begin{center}
\includegraphics[width=7cm]{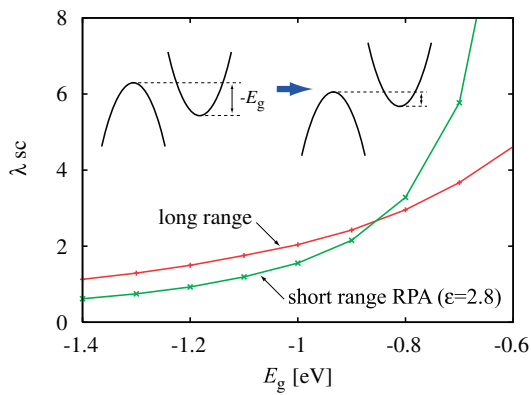}
\end{center}
\caption{
(Color online) $\lambda_{\rm sc}$ as a function of the band hybridization, $\Eg$, obtained with the long range $V(\bq)$ and by the RPA with $V_0 (\epsilon=2.8)$ for $\mu=0.3$ eV.
}
\label{Eg}
\end{figure}
%=======================================================================
	
	Finally, we give a prediction for further increase of $\Tc$.
	Within the unscreening mechanism discussed above, we can earn attractive interactions by decreasing $N(\EF)$, which have been carried out by changing $\mu$ in the previous paragraphs.
	%
	%This can be done by changing $\mu$, and there is another way: by changing the hybridization between conduction and electron band, i.e., $\Eg$.
	%
	There is another way to control $N(\EF)$: $N(\EF)$ can be reduced by decreasing the hybridization between conduction- and valence-band, i.e., $\Eg$.
	Actually, $\lambda_{\rm sc}$'s obtained both by the first correction $\Gamma^{(1)}(\bq)$ and the RPA with $V_0 (\epsilon=2.8)$ exhibit the tendency that $\Tc$ increase by decreasing $\Eg$ as shown in Fig. \ref{Eg} ($\mu=0.3$ eV).
	(Note that the $\bq$-dependence is relevant for small $|\Eg|$, so that the difference between the results with long range $V(\bq)$ and short range $V_0$ become large.
	At any rate, it is certain that $\lambda_{\rm sc}$ increases with decreasing $|\Eg|$.)
	At the present stage, a practical way to control $\Eg$ is unknown, but it could be actualized by, e.g., applying (chemical) pressure\cite{Takahashi08}.
	%

%%%%%%%%%%%%%%%%%%%%%%%%%%%%%%%%%%%%%%%%%%%%%%%%%%%%%%%%%%%%%%%%%%%%%%%%
%\section{Conclusion}
%-----------------------------------------------------------------------
	
	In conclusion, we have proposed a new mechanism for Fe-pnictide superconductor --- the unscreening effect.
	The direct Coulomb interaction is strengthened when $N(\EF)$ decreases since the screening becomes ineffective.
	As a result, $\Tc$ increases with decreasing $N(\EF)$.
	This is opposite from the previous theories of superconductivity, where the unscreening effect is neglected.
	The gap exhibits the $s_{\pm}$-wave symmetry, where the full gap is opened on the Fermi surfaces and their signs are opposite between the conduction- and valence-band.
	The magnitude of the gap roughly obeys $\Delta_{\rm c}^2/\Delta_{\rm v}^2 \simeq N_{\rm v}(\EF)/N_{\rm c}(\EF)$.

	$\Tc$ takes its maximum when the hole pockets almost vanish ($\mu\sim0.3$ eV in the present model).
	On the other hand, the SDW instability is the largest at around $\mu=0$ (undoped), since the SDW is much sensitive to the nesting property than to the unscreening effects.
	Consequently, the optimal doping for the SC is away from that for the SDW, due to the different sensitiveness to the unscreening and nesting conditions.
	These results explain the mysterious experimental results of photoemission and the nuclear relaxation rate.
	%
	%The phase diagram for the superconductivity and SDW has been obtained within the RPA.
	%
	%The superconductivity is sensitive to the unscreening effect, whereas the SDW is not.
	%
	%The SDW phase is rapidly suppressed away from the undoped region, and then the superconducting phase appears.
	%
	%The transition temperature of superconductivity, $\Tc$, increases with electron doping until the hole carriers vanish.
	%
	$\Tc$ can be also increased by reducing the hybridization between conduction- and valence-band.
	Furthermore, the present results is valid not only for the Fe-pnictides but also for the semimetal superconductors in general, namely, our theory indicates that the semimetals can be candidates for ``high-$\Tc$" superconductors due to the unscreening effect.
	%

%%%%%%%%%%%%%%%%%%%%%%%%%%%%%%%%%%%%%%%%%%%%%%%%%%%%%%%%%%%%%%%%%%%%%%%%
%\section*{Acknowledgment}

	%The authors greatly appreciate Hiroshi Fukuyama for stimulating discussions and useful advice.
	%
	%Thanks are also due to K. Kubo and T. Momoi for useful discussions.
	%
	%We acknowledge stimulating discussions with Hiroshi Fukuyama, K. Kubo and T. Momoi.
	%
	%This work was financially supported by Grant-in-Aid for Scientific Research on Priority Areas No. 17071002, and Next Generation Supercomputing Project, Nanoscience Program from MEXT.
	%
	This work was supported by Grant-in-Aid for Scientific Research on Priority Areas No. 16076203.
	Y. F. is supported by JSPS Research Fellowships for Young Scientists.

%\appendix

%%%%%%%%%%%%%%%%%%%%%%%%%%%%%%%%%%%%%%%%%%%%%%%%%%%%%%%%%%%%%%%%%%%%%%%%%


\begin{thebibliography}{99} %% The number "99" means that this list has more than nine items.
	\bibitem{Kamihara08}%2008-1-9
	Y. Kamihara, T. Watanabe, M. Hirano, and H. Hosono: J. Am. Chem. Soc. {\bf 130} (2008) 3296.
	
	\bibitem{Takahashi08}%2008-2-26
	H. Takahashi, K. Igawa, K. Arii, Y. Kamihara, M. Hirano, and H. Hosono: Nature {\bf 453} (2008) 376.
	
	\bibitem{XHChen08}%2008-3-25
	X. H. Chen, T. Wu, G. Wu, R. H. Liu, H. Chen, and D. F. Fang: Nature {\bf 453} (2008) 761.
	
	\bibitem{GFChen08}%2008-4-15
	G. F. Chen, Z. Li, D. Wu, G. Li, W. Z. Hu, J. Dong, P. Zheng, J. L. Luo, and N. L. Wang: Phys. Rev. Lett. {\bf 100} (2008) 247002.
	
	\bibitem{Ren08}%2008-4-24
	Z.-A. Ren, W. Lu, J. Yang, W. Yi, X.-L. Shen, Z.-C. Li, G.-C. Che, X.-L. Xiao, L.-L. Sun, F. Zhou, Z.-X. Zhao: Chin. Phys. Lett. {\bf 25} (2008) 2215.
	
	%\bibitem{Wen08}%2008-3-14
	%H.-H. Wen, G. Mu, L. Fang, H. Yang, and X. Zhu: Europhys. Lett. {\bf 82} (2008) 17009.
	
	\bibitem{Cruz08}
	C. de la Cruz, Q. Huang, J. W. Lynn, J. Li, W. Ratcliff, J. L. Zarestky, H. A. Mook, G. F. Chen, J. L. Luo, N. L. Wang, and P. Dai: Nature {\bf 453} (2008) 899.
	
	%\bibitem{Qiu08}
	%Y. Qiu, M. Kofu, W. Bao, S.-H. Lee, Q. Huang, T. Yildirim, J. R. D. Copley, J. W. Lynn, T. Wu, G. Wu, and X. H. Chen: Phys. Rev. B {\bf 78} (2008) 052508.
	
	\bibitem{Singh08}%2008-3-4
	D. J. Singh and M.-H. Du: Phys. Rev. Lett. {\bf 100} (2008) 237003.
	
	\bibitem{Haule08}%2008-3-9
	K. Haule, J. H. Shim, and G. Kotliar: Phys. Rev. Lett. {\bf 100} (2008) 226402.
	
	\bibitem{Mazin08}%2008-3-18
	I. I. Mazin, D. J. Sign, M. D. Johannes, and M. H. Du: Phys. Rev. Lett. {\bf 100} (2008) 057003.
	
	\bibitem{Cao08}%2008-3-21
	C. Cao, P. J. Hirshfeld, and H.-P. Cheng: Phys. Rev. B {\bf 77} (2008) 220506(R).
	
	\bibitem{Kuroki08}%2008-3-27
	K. Kuroki, S. Onari, R. Arita, H. Usui, Y. Tanaka, H. Kontani, and H. Aoki: Phys. Rev. Lett. {\bf 101} (2008) 087004.
	
	\bibitem{Ishibashi08}%2008-3-28
	S. Ishibashi, K. Terakura, and H. Hosono: J. Phys. Soc. Jpn. {\bf 77} (2008) 053709.
	
	\bibitem{Yin08}%2008-4-21
	Z. P. Yin, S. Leb\`egue, M. J. Han, B. P. Neal, S. Y. Savrasov, and W. E. Pickett: Phys. Rev. Lett. {\bf 101} (2008) 047001.
	
	\bibitem{Ma08}%2008-6-10
	F. Ma and Z.-Y. Lu: Phys. Rev. B {\bf 78} (2008) 033111.
	
	\bibitem{Baskaran08}
	G. Baskaran: J. Phys. Soc. Jpn. {\bf 77} (2008) 113713.
	
	\bibitem{Lee08}
	P. A. Lee and X.-G. Wen: arXiv:0804.1739.
	
	\bibitem{Barzykin08}
	V. Barzykin and L. P. Corkov: arXiv:0806.1933.
	
	\bibitem{Nomura08}
	T. Nomura: arXiv:0807.1168.
	
	\bibitem{Yanagi08}
	Y. Yanagi, Y. Yamakawa, and Y. \=Ono: arXiv: 0809.3189.
	
	\bibitem{Ikeda08}
	H. Ikeda: arXiv:0810.1828.
	
	\bibitem{Malaeb08}%2008-6-24
	W. Malaeb, T. Yoshida, T. Kataoka, A. Fujimori, M. Kubota, K. Ono, H. Usui, K. Kuroki, R. Arita, H. Aoki, Y. Kamihara, M. Hirano, and H. Hosono: J. Phys. Soc. Jpn. {\bf 77} (2008) 093714.
	
	\bibitem{Nakai}%2008-4-30
	Y. Nakai, K. Ishida, Y. Kamihara, M. Hirano, and H. Hosono: J. Phys. Soc. Jpn. {\bf 77} (2008) 073701.
	
	\bibitem{Takada78}
	Y. Takada: J. Phys. Soc. Jpn. {\bf 45} (1978) 786.
	
	\bibitem{Ando82}
	T. Ando, A. B. Fowler, and F. Stern: Rev. Mod. Phys. {\bf 54} (1982) 437.
	
	\bibitem{Kozlov65}
	A. N. Kozlov and L. A. Makisimov: Sov. Phys. JETP {\bf 21} (1965) 790.
	
	\bibitem{Halperin68}
	B. I. Halperin and T. M. Rice: Solid State Phys. {\bf 21} (1968) 115; Rev. Mod. Phys. {\bf 40} (1968) 775.
	
	\bibitem{dielectric}
	The dielectric function of LaFeAsO is estimated from the first principle calculation\cite{Nakamura08}.
	Its mean value is about $\epsilon^{\rm mean}\sim $2-3.
	
	\bibitem{Nakamura08}
	K. Nakamura, R. Arita, and M. Imada: J. Phys. Soc. Jpn. {\bf 77} (2008) 093711.
	
	\bibitem{doping}
	Here we estimated the carrier number with the tight-binding dispersion proposed by Kuroki et al.\cite{Kuroki08}.
	The hole pockets vanish at $x\simeq 0.34$.
	
\end{thebibliography}
\end{document}